\documentclass[aps,pra,showpacs,showkeys,preprintnumbers,amssymb,amsfonts,floatfix,a4paper,twoside,twocolumn]{revtex4}
\usepackage{amsmath,graphicx,epsfig}
\usepackage{color}
\usepackage{graphicx}% Include figure files
\usepackage{dcolumn}% Align table columns on decimal point
\usepackage{bm}% bold math
\usepackage{soul}% strikethrough text
\usepackage{mathpazo}
\usepackage{rotating}
\usepackage{multirow}

\begin{document}
%\preprint{APS/123-QED}
\title{Photodissociation of trapped Rb$_2^+$: Implications for simultaneous trapping of atoms and molecular ions}
\author{S. Jyothi$^1$, Tridib Ray$^{1,2}$, Sourav Dutta$^1$, A.R. Allouche$^3$, Romain Vexiau$^4$, Olivier Dulieu$^4$ and S. A. Rangwala$^1$}
\affiliation{%
$^1$Raman Research Institute, C. V. Raman Avenue, Sadashivanagar, Bangalore 560080, India.\\
$^2$Light-Matter Interactions Unit, Okinawa Institute of Science and Technology Graduate University, Onna, Okinawa 904-495, Japan.\\
$^3$Institut Lumi\`ere Mati\`ere, UMR5306 Universit\'e Lyon 1 - CNRS, Universit\'e de Lyon, 69622 Villeurbanne Cedex, France.\\
$^4$Laboratoire Aim\'e Cotton, CNRS, Universit\'e Paris-Sud, ENS Cachan, Universit\'e Paris-Saclay, Orsay Cedex, France.}
%%%%%%%%%%%%%%%%%%%%%%%%%%%%%%%%%%%%%%%%%%%%%%%%%%%%%%%%%%%%%%%%%%%%%%%%%%%%%%%%%%%%%%%%%%%%%%%%%%%%%%%
\begin{abstract}
The direct photodissociation of trapped $^{85}$Rb$_2^+$ (rubidium) molecular ions by the cooling light for the $^{85}$Rb magneto-optical trap (MOT) is studied, both experimentally and theoretically. Vibrationally excited Rb$_{2}^{+}$ ions are created by photoionization of Rb$_{2}$ molecules formed photoassociatively in the Rb MOT and are trapped in a modified spherical Paul trap. The decay rate of the trapped Rb$_{2}^{+}$ ion signal in the presence of the MOT cooling light is measured and agreement with our calculated rates for molecular ion photodissociation is observed. The photodissociation mechanism due to the MOT light is expected to be active and therefore universal for all homonuclear diatomic alkali metal molecular ions.   
\end{abstract}
%\keywords{Cold atoms, Hybrid trap, cold molecules etc} 
\maketitle
%%%%%%%%%%%%%%%%%%%%%%%%%%%%%%%%%%%%%%%%%%%%%%%%%%%%%%%%%%%%%%%%%%%%%%%%%%%%%%%%%%%%%%%%%%%%%%%%%%%%%%%
The spatially overlapped trapping of cold atoms and ions~\cite{Smith2005,Zipkes2010,Hall2011,Ravi2012a,ray2014,Grier2009,Hudson2011} has significantly expanded our ability to study interactions in cold, dilute gas ensembles. In particular, atomic ion-atom collisions, charge exchange collisions~\cite{Grier2009,Lee2013,Hudson2011,Mukaiyama2015,Sivarajah2012}, sympathetic cooling of ions by ultracold trapped atoms~\cite{Ravi2012b,Hall2012,Sivarajah2012,ray2014,sourav2015}, three body reactions~\cite{Arne2012,Denschlag2013b,Denschlag2016} and molecular ion formation processes~\cite{Hall2011,Humberto2015} have been investigated. Two complementary directions motivate key goals for future work, (a) the low partial wave ion-atom collisions which explores quantum scattering and many particle physics and (b), the controlled collisions between the cold molecular ions produced in the ion-atom traps with co-trapped neutral atoms~\cite{Rellergert2013} and with light.
\par
%%%%%%%%%%%%%%%%%%
 A critical question which arises is whether the molecular ions can be trapped for a substantial extent of time simultaneously with an ensemble of cold atoms in order to study the interaction between them. In this letter, we address the possibility of simultaneous trapping of $^{85}$Rb$_2^+$ molecular ions with ultracold $^{85}$Rb atoms in a magneto-optical trap (MOT). Our experimental observation shows that the cooling light for the Rb MOT leads to rapid destruction of the measured Rb$_2^+$ ion signal. We measure the lifetime of trapped Rb$_2^+$ in the presence of 780.2413 nm ($\equiv$ 12816.54 cm$^{-1}$) light to be 495$\pm$80 ms. We discuss possible dissociation mechanisms and show that the experimental observation is in agreement with our theoretical calculations for the photodissociation of Rb$_2^+$ molecular ions.The observed photodissociation mechanism is expected to be universal to all diatomic homonuclear alkali molecular ions as they exhibit similar potential energy characteristics.\par
\begin{figure}[ht]
\centering
\includegraphics[width=8.6 cm]{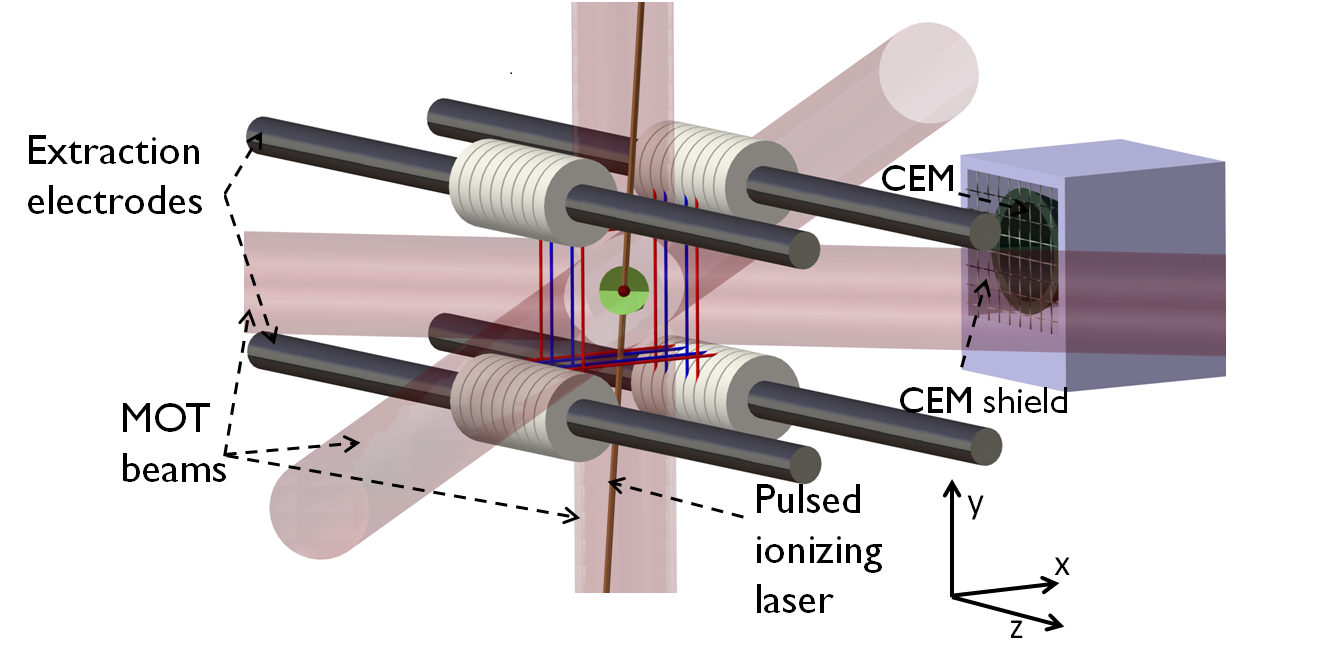}
\caption{(Color online) Schematic of the ion-atom hybrid trap, along with a channel electron multiplier for ion detection. The outer (red) wires are the ion trap end cap electrodes and the inner (blue) wires are the rf electrodes for the ion trap. The red sphere and the green cut sphere represent the spatial extent of the MOT and the modified spherical Paul trap respectively.
}
\label{fig:chamber}
\end{figure}
\begin{figure}[ht]
\centering
\includegraphics[width=8.6 cm]{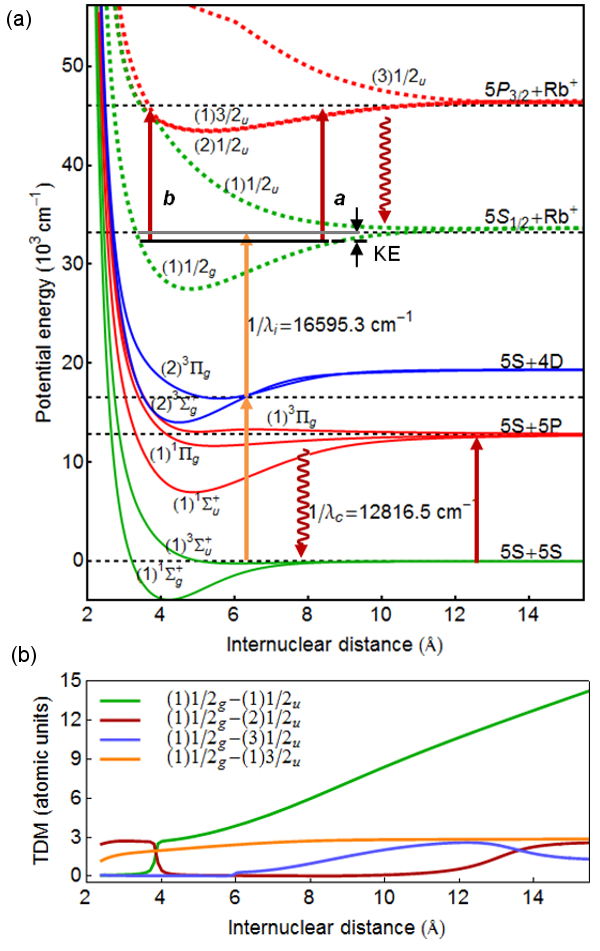}
\caption{(Color online) Selected potential energy curves (PECs) of Rb$_{2}$ and Rb$^{+}_{2}$ molecules relevant for the experiment are shown in panel (a). The Rb$^{+}_{2}$ PECs include spin-orbit interaction. Rb$_{2}$ molecules are formed using 12816.54 cm$^{-1}$ photons (red arrow) and are ionized using two photons of 16595.3 cm$^{-1}$ (orange arrows) to form Rb$^{+}_{2}$ molecules. The red arrows marked as \textit{a} and \textit{b} corresponds to the dissociation channels (a) and (b) described in the text. Panel (b) shows the selected transition electric dipole moments (TDM) for transitions from $(1) 1/2_g$ electronic state of Rb$_2^+$. The data for the plots is available in the supplementary material.}
\label{fig:IonCreation}
\end{figure}
The experimental system consists of an ion-atom hybrid trap assembly as shown in Fig.~\ref{fig:chamber}. The detailed description of the experimental system can be found in earlier work~\cite{Ray2013a,ray2014,Jyothi2015}. Briefly, the hybrid trap consists of a MOT for atoms and a modified spherical Paul trap made of four tungsten wire loops in a square shape geometry for trapping ions. The ion trap radio frequency (rf) of 500 kHz with 150 V amplitude is applied to the inner pair of wires and a small (-5 V) constant potential is applied on the outer wires. The ion trap is operated at the optimal trapping voltage for Rb$_2^+$ ions.\par
%%%%%%%%%%%%%%%%%%%%%%%%%%%%%%%%%%%%%%%%%%%%%%%%%%%%%%%%%%%%%%%%%%%%%%%%%%%%%%%%%
For the experiment, $^{85}$Rb atoms are cooled and trapped in the MOT using three pairs of mutually orthogonal counter-propagating laser beams of $\sim 8$ mm diameter intersecting at the center of a gradient magnetic field. The cooling laser beam is red detuned from the $5S_{1/2}$~(F=3)~$\leftrightarrow$~5$P_{3/2}$~$(F^{\prime}$=4) transition by 12 MHz. The repumper light is on resonance with the $5S_{1/2}$~(F=2)~$\leftrightarrow$~5$P_{3/2}$~$(F^{\prime}$=3) transition. Approximately 40 mW of cooling and 3 mW of repumper power is used and is equally distributed over the six laser beams.\par
%%%%%%%%%%%%%%%%%%%%%%%%%%%%%%%%%%%%%%%%%%%%%%%%%%%%%%%%%%%%%%%%%%%%%%%%%%%%%%%%%%%%%%%%%%%%%%%%%%%
The Rb$^{+}_{2}$ molecular ions are created by ionizing neutral Rb$_2$ molecules produced by photoassociation in the MOT~\cite{Gabbanini2000,Caires2005}. The potential energy curves (PEC) of Rb$_{2}$~\cite{Allouche2012} and Rb$^{+}_{2}$~\cite{AllouchePC} molecules relevant for this experiment are shown in Fig.~\ref{fig:IonCreation}(a). Two Rb atoms in the ground state photoassociate in the presence of a cooling laser photon $(1/\lambda_{c}= 12816.54$ cm$^{-1})$ to form a loosely bound molecule in the excited electronic state. The excited molecule spontaneously decays either to a highly vibrationally excited bound molecule in the electronic ground states ($^3\Sigma_u^+$ or $^1\Sigma_g^+$) or to two free ground state atoms by emitting a photon. These loosely bound neutral Rb$_{2}$ molecules are ionized by two photons of 602.5 nm $(2/\lambda_{i}= 33190.6$ cm$^{-1}$) to produce Rb$_2^+$ ions in their electronic ground state $(1)1/2_g$ (or $X^2\Sigma_g^+$ in Hund's case "a" notation) as previously shown by Gabbanini et. al.~\cite{Gabbanini2000}. The ionization laser is a pulsed dye laser, pumped by the second harmonic of a Nd:YAG laser (10 Hz repetition rate).\par
%%%%%%%%%%%%%%%%%%%%%%%%%%%%%%%%%%%%%%%%%%%%%%%%%%%%%%%%%%%%%%%%%%%%%%%%%%%%%%%%%%%%%%%%%%%%%%%
%%%%%%%%%%%%%%%%%%%%%%%%%%%%%%%%%%%%%%%%%%%%%%%%%%%%%%%%%%%%%%%%%%%%%%%%%%%%%%%%%%%%%%%%%%%%%%%
Energy consideration allows only those vibrational levels of Rb$_2^+$ which have binding energies $\geq$500.2 cm$^{-1}$ (= Ionization potential(Rb) 33690.8 cm$^{-1}$ - 2$/\lambda_{i}$) to be populated. Based on the calculation of vibrational energies~\cite{LeRoy2007} using the ab initio potentials~\cite{AllouchePC}, the vibrational level with binding energy close to 500.2 cm$^{-1}$ is $v = $174 of $(1)1/2_g$ electronic state, which essentially implies that the ionization process can create Rb$_2^+$ in levels $v \leq$ 174. The kinetic energy (KE) of the electron determines the initial vibrational levels in which the Rb$_2^+$ ions are created. 

\begin{figure}[b]
\centering
\includegraphics[width=8 cm]{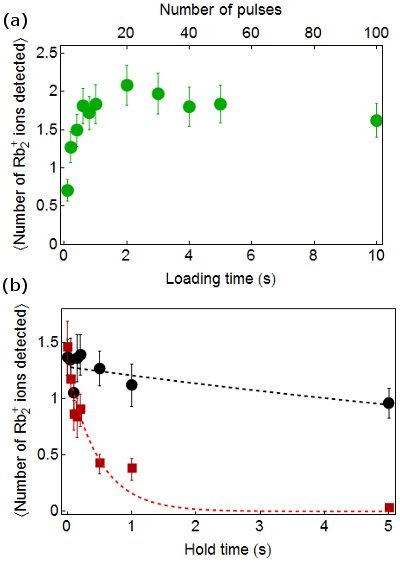}
\caption{(Color online) Panel (a) shows the mean number of Rb$^+_2$ ions detected as a function of loading time (the top axis corresponds to the number of ionization pulses). The ion number initially increases and then saturates. The MOT trapping lasers are kept \textit{on} during the experiment. Panel (b) shows the mean number of Rb$^{+}_{2}$ ions detected, for different hold times. The black circles represent the lifetime in the absence of MOT light fields and the red squares shows the lifetime in the presence of cooling laser field. The distribution of number of ions detected for each hold time is presented in the supplementary material. The error bars are the standard deviation of the mean of the bootstrapped data sets. [see supplementary material].
}
\label{fig:Rb2+_All}
\end{figure}
Since the molecular ions are created from the MOT, they are created at the centre of the ion trap. The ion trap voltages are $\textit{on}$ during the ionization process so that the ions are trapped as they are produced. The trapped ions are detected by extracting them onto a channel electron multiplier (CEM), by switching the voltages appropriately on a set of electrodes \cite{ray2014}. Prior to the extraction, the trapping rf field is switched \textit{off} to mass separate any Rb$^+$ ions from Rb$_2^+$ ions~\cite{Jyothi2015}.\par
The process of loading of Rb$_2^+$ into the ion trap is monitored by extracting the ions at different times during the loading process as shown in Fig.~\ref{fig:Rb2+_All}(a). In our experiment, the molecular ions created per shot fluctuate due to the energy fluctuation of the ionizing dye laser pulse. For this reason, we repeated the experiment at each hold time more than 40 times. A steady state is reached when the ion production rate equals the ion loss rate and with a loading time of 10s, the ion trap has 1.62$\pm$0.22 Rb$_2^+$ ions. In principle, the ion loss rate depends on factors such as rf heating, collisions with background gases, collisions with ultracold atoms and on the presence of light. In order to determine the dominant loss channel, we load the trap to steady state and hold the ions in the ion trap for variable hold time in different scenarios followed by ion extraction onto the CEM to count the number of ions survived.\par
The effect of Rb D$_2$ light on the population of Rb$^{+}_{2}$ molecular ions is studied by measuring the lifetime of Rb$^{+}_{2}$ when held in the presence of MOT cooling light. Each experimental cycle consists of the following sequence: loading the MOT to steady state, turning on the ion trap, turning on the pulsed dye laser to create Rb$^{+}_{2}$, turning off the pulsed dye laser (thus stopping further creation of Rb$^{+}_{2}$), removing the MOT atoms by blocking the repumper light and then holding the Rb$^{+}_{2}$ ions in the ion trap for a predetermined hold time(t) either in presence or absence of the MOT cooling light. At the end of the hold time, the ions are extracted out of the ion trap and detected by the CEM. The hold time is varied and the sequence is repeated. Fig.~\ref{fig:Rb2+_All}(b) shows the mean number of Rb$^{+}_{2}$ ions detected, for different ion trap hold times either in presence (red squares) or absence (black circles) of the MOT cooling light. It should be mentioned that, during the Rb$^{+}_{2}$ ion creation process a large number of Rb$^{+}$ ions are also created. However the ion trap parameters favor molecular ions and atomic ions escape from the ion trap within a few tens of milliseconds.To avoid systematic effects due the presence of atomic ions, we restrict the our analysis to t$\geq$50 ms. We fit the average number of Rb$^{+}_{2}$ ions to A exp(-t/$\tau$) to obtain the lifetime of Rb$^{+}_{2}$ in the ion trap. In the absence of MOT lights the lifetime is 16$\pm$8 s and is dramatically reduced to 495$\pm$80 ms in the presence of the cooling light. The single exponential fit allows a comparison between the with and without light decay. While different vibrational levels decay at different rates  in the presence of MOT light (see below), the resolution of the present experiment is not enough to distinguish between them, and so the single exponential is representative of a rate for the disintegration process.\par
%%%%%%%%%%%%%%%%%%%%%%%%%%%%%%%%%%%%%%%%%%%%%%%%%%%%%%%%%%%%%%%%%%%%%%%%%%%%%%%%%%%%%%%%%%%%%%%%%%%%%%%%%%%%%%%%%%%%%%%%%%%%%%
We consider two possible channels for the disintegration of the Rb$^{+}_{2}$ ground state molecules induced by the MOT cooling light. The dipole allowed transitions from the initial $(1)1/2_g$ state lead to the states $(1)1/2_u$, $(2)1/2_u$, $(3)1/2_u$ and $(1)3/2_u$ (written in Hund's case c notation including spin-orbit interaction), with the relevant electric transition dipole moment (TDM)~\cite{AllouchePC} shown in Fig.~\ref{fig:IonCreation}(b). Of these states, the $(3)1/2_u$ is energetically not accessible. The $(1)1/2_u$ state is of $(1)^2\Sigma_u^+$ character below 4 $\AA$, and changes into  $(1)^2\Pi_u$ before 4 $\AA$ due to an avoided crossing induced by the spin-orbit interaction. This explains the crossing of TDM curves in Fig.~\ref{fig:IonCreation}(b) around 4 $\AA$. The possible dissociation channels are shown in Fig.~\ref{fig:IonCreation}(a) and are:\newline
(a) bound-to-bound excitation (indicated by \textit{a} in Fig.~\ref{fig:IonCreation}(a)) followed by spontaneous decay to ion-atom pair.\newline
Rb$^{+}_{2}+h\nu_{c}\rightarrow $(Rb$^{+}_{2})^*\rightarrow~$Rb~+~Rb$^++h\nu$ \newline
(b) direct photodissociation (PD) from $(1) 1/2_g$ to $(1) 1/2_u$ (indicated by \textit{b} in Fig.~\ref{fig:IonCreation}(a)). \newline
Rb$^{+}_{2}+h\nu_{c}\rightarrow $~Rb~+~Rb$^+$ \newline
The possibility for the dissociation of Rb$^{+}_{2}$ ions through these excitation channels are investigated by calculating the transition rates.\\
\begin{figure}[b]
\centering
\includegraphics[width=8.4 cm]{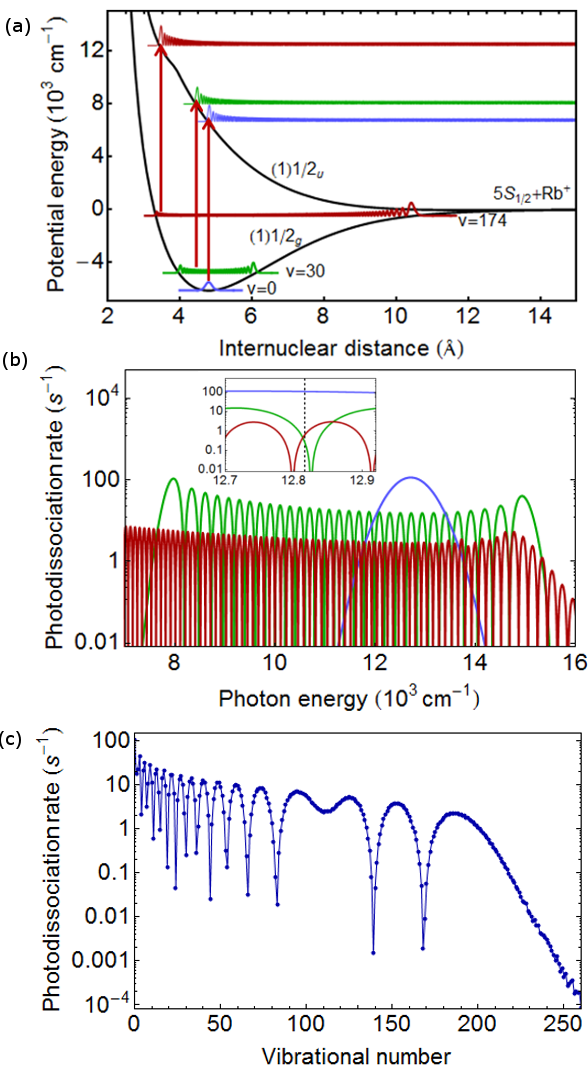}
\caption{(Color online) Panel (a) shows the $(1)1/2_g$ and $(1)1/2_u$ PECs, the wavefunctions for selected vibrational levels $v=$ 0 (blue), 30 (green) and 174 (red) and the final unbound wave functions at the cooling light photon energy. Panel (b) shows the corresponding photodissociation rates as a function of the photon energy. The inset shows a zoom at energies close to $h\nu_{c}$, indicated by the vertical dashed line. In panel (c), the calculated photodissociation rate at $h\nu_{c}$ is shown for all the vibrational levels of the $(1)1/2_g$ state.}
\label{fig:PD2}
\end{figure}
In the case \textit{a}, the energy of the cooling light $h\nu_{c}$, where $\nu_c$ is the frequency, is insufficient to excite the molecule to the continuum of the first excited electronic state (dissociation to $5P_{1/2}+$Rb$^+$ asymptote). However, one can consider a resonant bound to bound transition and subsequent de-excitation to the ground state. If the de-excitation to the ground state continuum dominates over that to the bound levels, this process can be considered as a dissociation channel. We first calculate the bound-to-bound excitation by the MOT cooling light from various possible initial vibrational levels of the $(1)1/2_g$ state to the $(2)1/2_u$ and $(1)3/2_u$ states. The accuracy of the ab initio PECs is not high enough to predict the vibrational levels exactly and it is thus not possible to say if the excitation is resonant or off-resonant, although the latter is more likely. Assuming resonant excitation to the nearest vibrational level determined from photon energy consideration, the maximum excitation rate is calculated to be of the order of MHz or smaller for the $(1)1/2_g$ $\rightarrow$ $(1)3/2_u$ and $(1)1/2_g$ $\rightarrow$ $(2)1/2_u$ transitions. The spontaneous emission rate to the bound and free states of $(1)1/2_g$ is then calculated for both the $(2)1/2_u$ and $(1)3/2_u$ states. For both electronic states we find that the bound-to-free spontaneous emission rate is negligibly small compared to the bound-to-bound spontaneous emission rate, eliminating the possibility of this dissociation channel.\\
%%%%%%%%%%%%%%%%%%%%%%%%%%%%%%%%%%%%%%%%%%%%%%%%%%%%%%%%%%%%%%%%%%%%%%%%%%%%%%%%%%%%%%%%%%%%%%%%%%%%%%%%%%%%%%%%%%%%%%%%%%%%%%
%
In the direct photodissociation process (case b), a bound Rb$_2^+$ molecule in $(1)1/2_g$ absorbs a photon $h\nu_{c}$ (the slower rate determining step) to reach the $(1)1/2_u$ state and dissociates immediately to the continuum of $(1)1/2_u$ to form Rb and Rb$^+$ as shown in Fig.~\ref{fig:PD2}(a). The photodissociation cross section $\sigma_{pd}$ is proportional to the photon energy and the Franck-Condon overlap between the levels involved~\cite{LefebvreField,Gordon1969}. The rate of photodissociation is given by
\begin{align}
R_{pd}&=\sigma_{pd} \times \mathcal{F}\\
%R_{pd}&=2.69\times10^{-18} \Delta E\left|\left\langle \psi_E(r)|d(r)|\psi_v(r)\right\rangle\right|^2\frac{I}{h\nu_{c}}
&=\frac{4\pi^2a_0^2}{3 \hbar c} \Delta E\left|\left\langle \psi_E(r)|d(r)|\psi_v(r)\right\rangle\right|^2\frac{I}{h\nu_{c}}
\label{eq:PDrate}
\end{align}
where $\sigma_{pd}$ is the dissociation cross section, $\mathcal{F}=I/h\nu_c$ is the dissociating light flux, \textit{I} is the intensity of the light, $a_0$ is the Bohr radius, 2$\pi\hbar=h$ is the Planck's constant, c is the velocity of light, $\Delta E$ is the photon energy and \textit{d(r)} is the transition dipole moment between $(1)1/2_g$ and $(1)1/2_u$ states~\cite{AllouchePC}. The continuum wavefunction $\psi_E(r)$ is calculated using Cooley-Numerov method~\cite{Levine2009} and the bound wavefunction $\psi_v(r)$ is calculated using LEVEL code~\cite{LeRoy2007}.\par
The variations of the calculated PD rates vs the photon energy reproduce the oscillations of the initial bound wave functions, as shown for few vibrational levels in Fig.~\ref{fig:PD2}(b)~\cite{Blange1996,Gordon1969}. Fig.~\ref{fig:PD2}(c) shows the PD rate (at 12816.5 cm$^{-1}$ with 0.16 W/cm$^2$ of intensity) for all the vibrational levels of Rb$_2^+$ ground state. The results show that molecular ions in different vibrational levels dissociate at different rates in presence of the MOT cooling light with the trend that the lower vibrational levels dissociate at faster rates.\par
In the experiment, we cannot determine precisely the vibrational level of the initially created Rb$_2^+$ because the details of the Rb$_2$ photoionization process are unknown and difficult to compute. The two limits concerning the vibrational level of the initially created Rb$_2^+$ are determined by the KE of the ejected electron. (i) In case the electron is ejected with zero KE, the Rb$_2^+$ ions are created in/near $v=$174 which dissociates at the calculated rate of 0.7 s$^{-1}$. However, we note that the dissociation rate for v = 174 could be in the range 0.7 to 2.1 s$^{-1}$, considering the uncertainty in the dissociation energy of the ab initio PEC compared to the experimental dissociation energy~\cite{Bellos2013}, which is higher by 120 cm$^{-1}$ compared to theory, and causes shift of the calculated vibrational levels and hence the PD rates. (ii) In case of electron ejection with all possible KE, all vibrational levels of Rb$_2^+$ from $v=$ 0 to 174 may be populated. An average of calculated PD rates (Fig.~\ref{fig:PD2}(c)) over all vibrational levels below $v=$ 174 yields 5.9 s$^{-1}$ but the actual rate could be anywhere between 0.001 to 100 s$^{-1}$ as shown in Fig. 4(c).The experimentally measured rate 2.0$\pm$0.3 s$^{-1}$is consistent with either of the above cases. Further detailed experimental and theoretical work is thus needed to establish the initial vibrational distribution of Rb$_2^+$. However, even in the absence of such information, the experimental and theoretical results presented above firmly establish that light near D$_2$ transition induces dissociation of Rb$_2^+$. It should also be noted that the commonly used 1064 nm light for optical dipole trapping would dissociate loosely bound Rb$_2^+$ ($v>$10) by a similar process but have no significant effect on deeply bound Rb$_2^+$ (e.g. $v=$0). \par
The above experiments and calculations show that the direct photodissociation of Rb$^+_2$ ion by the cooling light of the MOT is the dominant mechanism for the loss of trapped Rb$^+_2$ molecular ions in all vibrational states. Since all homonuclear alkali (X) molecular ion dimers, X$_2^+$, have a similar arrangement of their molecular PECs~\cite{Sharp1971}, we conclude through preliminary inspection that this photodissociation process will be active and influential for all such molecular ion systems. This would imply that in the presence of the MOT cooling light, long lived trapping of X$_2^+$ will be challenging, and creating sizable steady state ensembles of X$_2^+$ ions and parent alkali atoms would be experimentally difficult. A possible solution to the problem is the use of a far-detuned optical dipole trap for ultracold atoms instead of a MOT. \par
%%%%%%%%%%%%%%%%%%%%%%%%%%%%%%%%%%%%%%%%%%%%%%%%%%%%%%%%%%%%%%%%%%%%%%%%%%%%%%%%%%%%%%%%%%%%%%%%
S. J. and S. A. R. acknowledge Krishna Rai Dastidar for assistance in the setting up of the molecular level calculations. S. D. acknowledges support in the form a Pancharatnam Fellowship from RRI. A. R. A. acknowledges the access to the HPC resources of the FLMSN,"F\'ederation Lyonnaise de Mod\'elisation et Sciences Numeriques", partner of EQUIPEX EQUIP@MESO and  HPC of the "Centre de calcul CC-IN2P3" at Villeurbanne. O. D. and S. A. R. acknowledge support from the Indo-French Centre for the promotion of Advanced Research-CEFIPRA project 5404-1.
%\bibliographystyle{plain}
%\bibliography{reference}

\end{document}